\documentclass[fleqn,usenatbib]{mnras}

\usepackage{newtxtext,newtxmath}

\usepackage[T1]{fontenc}

\DeclareRobustCommand{\VAN}[3]{#2}
\let\VANthebibliography\thebibliography
\def\thebibliography{\DeclareRobustCommand{\VAN}[3]{##3}\VANthebibliography}

\usepackage{graphicx}	
\usepackage{amsmath}	

\title[Shubham singh et al. 2023]{Single pulse polarization study of pulsars B0950+08 and B1642-03: micropulse properties and mixing of orthogonal modes}

\author[Singh et al.]{
Shubham Singh,$^{1}$\thanks{E-mail: ssingh@ncra.tifr.res.in}
Yashwant Gupta,$^{1}$
Kishalay De$^{2,3}$
\\
$^{1}$National Centre for Radio Astrophysics, Tata
Institute of Fundamental Research, Pune 411 007, India\\
$^{2}$MIT-Kavli Institute for Astrophysics and Space Research, 77 Massachusetts Ave., Cambridge, MA 02139, USA\\
$^{3}$NASA Einstein Fellow\\}

\date{Accepted XXX. Received YYY; in original form ZZZ}

\pubyear{2022}

\begin{document}
\label{firstpage}
\pagerange{\pageref{firstpage}--\pageref{lastpage}}
\maketitle

\begin{abstract}
 We present the results of a high-time resolution polarization study of single pulses from pulsars B0950+08 and B1642-03. Single pulses from pulsar B0950+08 sometimes show isolated micropulses without any significant associated subpulse emission. Assuming that the properties of such micropulses represent the intrinsic nature of micropulse emission, we characterize the width and polarization properties of these `intrinsic' microstructures. Most of the `intrinsic' micropulses ($\sim 90\%$) follow common characteristic polarization properties, while the average width of these micropulses is consistent with the general micropulse population from this pulsar. Single pulses from these pulsars show a diverse range of polarization properties, including depolarization and mixing of two orthogonal modes resulting in polarization position angle jumps. We present a superposition model of the two orthogonal modes which can explain depolarization, the observed position angle jumps, and associated changes in other polarization parameters.
\end{abstract}




\section{Introduction}
Mechanism giving rise to radio emission from pulsars is still not fully understood (e.g. \citet{Beskin_2018}, \citet{Mitra_2017}, and \citet{Cerutti_2016}). Observational properties of pulsars provide clues about the origin of the radio emission. While folded profile gives information about the average emission pattern across the radio beam, single pulse phenomena like drifting, nulling, and mode changing provide insight into the exact process of emission. Polarization properties of folded profiles and individual single pulses carry valuable information about the geometry of the pulsar, the emission pattern, and the process of emission. Radio emission from pulsars is known to be significantly linearly and circularly polarized. Similar to the total intensity profile, folding Stokes data of thousands of single pulses generates a stable folded profile for each of the Stokes parameters. The polarization position angle (PPA) of the polarization profile of many pulsars traces an S-shaped curve (\citet{MSPES} and \citet{Hankins_2009}). This S-shaped trace of PPA is explained by the Rotating Vector Model (RVM, \citet{RVM}). The S-curve trace of PPA is also used to determine the geometry of pulsars and the radio emission heights \citep{Mitra_2017}.\\

Unlike folded profiles, individual single pulses from pulsars show unstable polarization properties, which change from pulse to pulse. Both the linear and circular polarization fractions vary largely for single pulses. Sometimes, parts of single pulses show close to $100\%$ polarization. According to current understanding, single pulses with close to $100\%$ polarization are only possible if radiation is produced by coherent curvature radiation \citep{minhajur}. The depolarization seen in single pulses is often attributed to superposition of the two orthogonal modes (\citet{stineberg1}, \citet{stineberg2} and \citet{gangadhar}). Different models of superposition of the two orthogonal modes have been often used to explain the complex polarization properties of single pulses (\citet{Primak_2022}, \cite{dykes_2018}, \cite{gangadhar}, and \cite{stineberg1}). The presence of circular polarization in pulsar radiation is sometimes explained as a result of coherent superposition of two orthogonal modes \citep{cheng_1979}. Few other observational evidences of coherent superposition of the two modes are presented in  \citet{dykes_2018}. \\

The single pulses are often made of one or more components called subpulses. Polarization properties (polarization fractions and position angle) can vary across a subpulse window. Some pulsars also show narrower quasi-periodic structures on top of these subpulses  \citep{craft_1968}. These narrower structures are called `microstructures'. In this study, we focus on two bright pulsars, pulsar B0950+08 \citep{B0950_discovery} and pulsar B1642-03 \citep{B1642_discovery}, that show clear microstructures in their single pulses. We study the microstrcutres and polarization properties of single pulses from these two pulsars using the Giant Meterwave Radio Telescope (GMRT, \citet{GMRT_govind}). The study of small timescale structures in single pulses requires a sensitive high-time resolution observation. The GMRT is one of the most sensitive radio telescopes at lower frequencies. With the GMRT, we can obtain a high-quality, low-frequency single pulse observation of these two pulsars.\\   

pulsar B0950+08 is the first pulsar in which microstructures were identified \citep{craft_1968}. Since then, microstructures from this pulsar have been extensively studied (\citet{kuzmin}, \citet{popv_2002}, \cite{hankin_nature}, and \citet{Mitra_2015}). A wide range of microstructure widths has been reported from this pulsar starting from 0.8$\mu$s \citep{hankin_nature} to a few hundreds of $\mu$s \citep{popv_2002}. Though, sub-microsecond structures were not found in a higher time resolution study \citep{popv_2002}, still structures with time scales of $\sim7~\mu s$ were reported by \citet{popv_2002}. The average timescale of microstructures of this pulsar is $\sim$ 100 $\mu$s (\citet{cordes_1990} and \citet{popv_2002}), which is consistent with the microstructure timescale relation with period of pulsar (\citet{Cordes_79}, \citet{kramer_2002}, and \citet{kishlay}). There has been no study reporting the timescales of microstructures from pulsar B1642-03 to the best of our knowledge.\\

There are only a few studies reporting polarization properties of microstructures (e.g. \citet{cordes_hankin_77}, \citet{cordes_polarization}, \citet{Mitra_2015}). One of the difficulties in such studies is separating microstructures from subpulse emission. One useful way is to subtract an approximated smooth subpulse to separate out the micropulses riding on top of the subpulse emission \citep{Mitra_2015}. While this method is useful in estimating the micropulse widths and periodicity, estimates of linear and circular polarization fractions obtained from such analysis are not accurate as we are trying to separate out polarization properties of micropulses from subpulses, without knowing how the polarized part of the subpulse and micropulse were combined. There are other complications in the polarization properties of the single pulses, such as depolarization and position angle jumps, that further complicate the situation. Also, there is no way to know the polarization position angle of microstructures in such an analysis. \citet{cordes_hankin_77} argue that the polarization properties of micropulses from the auto-correlation analysis of single pulses have some bias, as the contributions from individual micropulses in the auto-correlation function (ACF) are weighted by their strength and the results would be biased towards strong micropulses. These complications can be avoided by looking at individual microstructures that are not corrupted by subpulse emission. Such isolated microstructures are expected to represent the intrinsic nature of micropulses and can offer a unique opportunity to probe the intrinsic properties of microstructures.\\

In this work, we present the polarization properties of micropulse emission from pulsar B0950+08 and report the timescales of microstructures from pulsar B1642-03. We also try to understand the polarization properties of single pulses (especially depolarization and the position angle mode jumps) of pulsars B0950+08 and B1642-03, using the coherent and incoherent superposition of the two orthogonal position angle modes. In section \ref{sec:sec2} of this paper, we describe the details of observations and data reduction. We present the polarization profiles of these two pulsars in section \ref{sec:sec3}. Section \ref{sec:sec4} talks about the properties of micropulses from pulsar B0950+08. In section \ref{B1642_micro}, we present the study of microstructures from pulsar B1642-03. We describe the simple empirical models to explain the polarization properties of single pulses from pulsars B0950+08 and B1642-03 in section \ref{sec:sec5}. Then we summarize the work in section \ref{sec:summary}. 

\section{Observations and data reduction}\label{sec:sec2}

The polarization data of pulsars B0950+08 and B1642-03, used in this study, were recorded with the legacy GMRT system \citep{GMRT_govind}. We use voltage time series of two polarization channels (R and L) recorded at the Nyquist rate. Data were recorded with a bandwidth of 32 MHz in band-3 of GMRT, covering the frequency range 306-338 MHz with the GMRT Software backend (GSB, \citet{GSB}). We use an offline coherent dedispersion routine \citep{GSB_CDP} to dedisperse the voltage time series of the two circular polarization channels (R and L). After dedispersion, we generate a filterbank file with 128 channels for all 4 Stokes parameters. The Stokes parameters for each channel and time sample were calculated after correcting for the gain difference between R and L polarization channels. The final resolution of this filterbank file is 15.36 $\mu$s. A small delay between the R and L voltage time series results in a linear gradient in polarization position angle (PPA) as a function of frequency and reduction in the linear polarization fraction. A delay calibration is done by correcting for the linear gradient of PPA as a function of the frequency channel. This correction also takes care of the variations caused by the rotation measure (RM) of these pulsars. We ignore small non-linearity in PPA versus frequency plot caused by the rotation measure (RM), as the values of RM for the two target pulsars are small and won't have a dominant non-linear effect over the narrow bandwidth of 32 MHz.\\
\begin{table}
	\centering
	\caption{List of observations used in this study. The first two entries are observations with the legacy GMRT system and full stokes data at a resolution of 15.36 $\mu$s were generated from these observations. The observations in the last three entries were taken with the upgraded GMRT and contain only total intensity information. The last observation in the table is raw voltage data recorded at Nyquist rate in parallel with the wideband observation on 02 Feb 2021 and was used to generate a very high time resolution total intensity time series. Data of all these observations were coherently dedispersed, as described in the text.}
	\label{tab:observations}
	\begin{tabular}{lcccccr} 
		\hline
		Date & Pulsar & Backend & Duration & Frequency & Time\\
             & & &  (minutes) & -range & -res.\\
             & & & & (MHz) &  ($\mu$s)  \\
		\hline
            13/04/2016 & B0950+08 & GSB & 15 & 306-338 & 15.36 \\
            04/04/2016 & B1642-03 & GSB & 24 & 306-338 & 15.36 \\
		\hline
            01/12/2020 & B0950+08 & GWB & 40 & 300-500 & 5.12 \\
            02/01/2021 & B0950+08 & GWB & 60 & 550-750 & 5.12 \\
            02/01/2021 & B0950+08 & GSB & 60 & 651-684 & 0.015\\
            \hline
            
	\end{tabular}
\end{table}

After noticing interesting unresolved structures at a resolution of 15.36 $\mu$s in the single pulses from pulsar B0950+08, We followed it up with the upgraded GMRT (uGMRT, \citet{UGMRT}). We observed pulsar B0950+08 with a wide bandwidth of 200 MHz in both band-3 (300-500 MHz) and band-4 (550-750 MHz) using the GMRT Wideband Backend (GWB, \citet{GWB_GMRT}). We get the final resolution of 5.12 $\mu$s in the wideband observations. The wideband data was cleaned using gptool\footnote{https://github.com/chowdhuryaditya/gptool}, an RFI filtering software developed specifically for GMRT time series data. We also recorded raw voltage data with a bandwidth of 32 MHz in the background of these observations. We used an online coherent dedispersion pipeline (CDP) available in uGMRT to dedisperse the wideband data. Nyquist data recorded with narrower bandwidth along with wideband observations were used to generate total intensity time series with very high time resolutions. This data was also coherently dedispersed by the offline dedispersion routine and then the power of dedispersed R and L time series was added sample-wise to get the total intensity time series with ultra-high time resolution. We can get up to $\sim$15 ns time resolution using this data. Table \ref{tab:observations} summarizes all the observations used in this study.

\section{polarization properties of folded profiles of pulsars B0950+08 and B1642-03}\label{sec:sec3}
\begin{figure}
    \centering
    \includegraphics[width=\columnwidth]{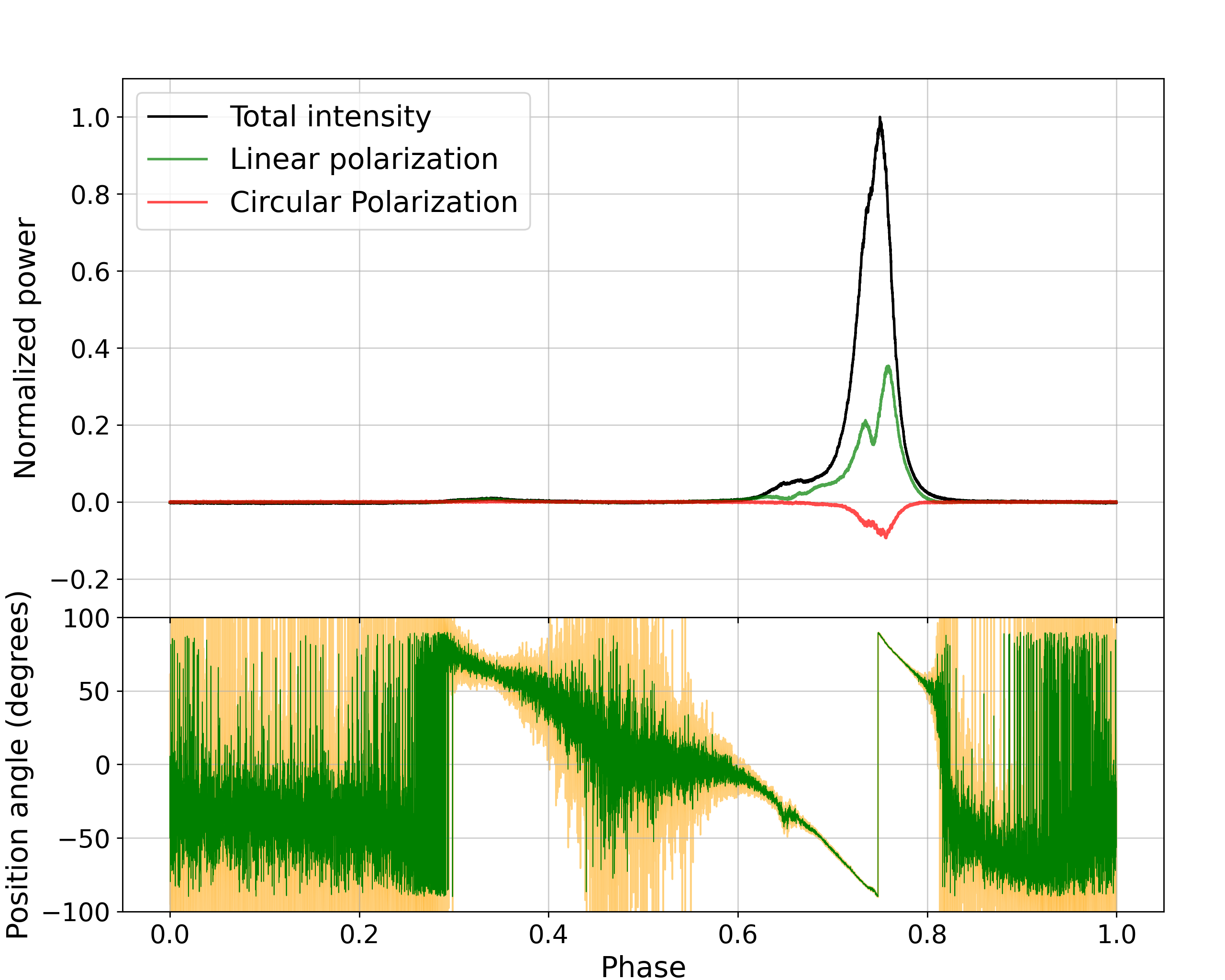}
    \caption{Folded profile of pulsar B0950+08 at 322 MHz (32 MHz bandwidth). The upper panel shows the folded profile in total intensity, linear polarization and circular polarization, while the lower panel shows the polarization position angle (PPA) versus the rotation phase of the pulsar. The plot is normalized by the peak intensity of the folded profile. The orange-shaded region in the lower panel represents the uncertainty in the position angle.}
    \label{fig:fig1}
\end{figure}

\begin{figure}
    \centering
    \includegraphics[width=\columnwidth]{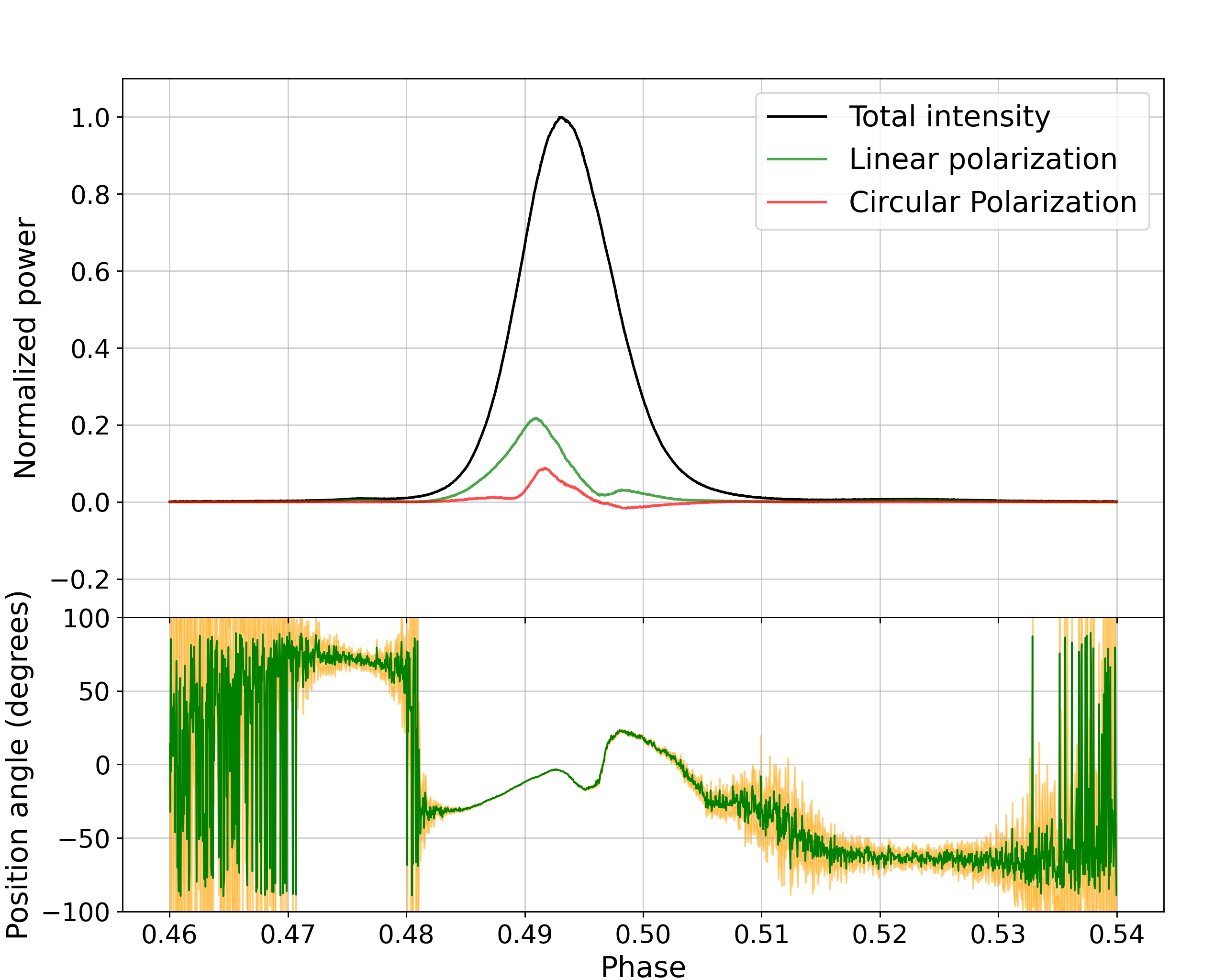}
    \caption{Folded profile of pulsar B1642-03 at 322 MHz (32 MHz bandwidth). The upper panel shows the folded profile in total intensity, linear polarization, and circular polarization, while the lower panel shows the polarization position angle (PPA) versus the rotation phase of the pulsar. The plot is zoomed around the pulse phase containing the signal from the pulsar and normalized by the peak intensity of the folded profile. The orange-shaded region in the polarization position angle plot represents the uncertainty in the position angle.}
    \label{fig:fig2}
\end{figure}

The voltage data recorded with the narrow band receiver of legacy GMRT was coherently dedispersed and the voltage series of the two circular polarizations were combined to get time series of Stokes parameters at a time resolution of 15.36 $\mu s$. We folded the time series at the topo-centric periods to generate folded profiles of the two pulsars with polarization parameters. Figure \ref{fig:fig1} shows the polarization profile of pulsar B0950+08 at a central frequency of 322 MHz. The lower panel of the plot shows a clear, smooth position angle track below the pulse profile. A presence of low level emission is evident in a large portion of the pulse phase from the polarization position angle (PPA) plot. Recently, \citet{B0950_fast} detected this pulsar over the whole pulse phase using the FAST telescope. Figure \ref{fig:fig2} shows the folded profile of pulsar B1642-03 at central frequency of 322 MHz and a bandwidth of 32 MHz. The position angle track shown in the lower panel is fairly complicated and does not show any obvious RVM track. Recently, \citet{B1642_mitra} have found RVM track in this pulsar by using only the highly polarized parts of the single pulses. The circular polarization changes sign almost at the middle of the profile.

\section{`Intrinsic' Micropulses from pulsar B0950+08}\label{sec:sec4}

\begin{figure}
    \centering
    \includegraphics[width=\columnwidth]{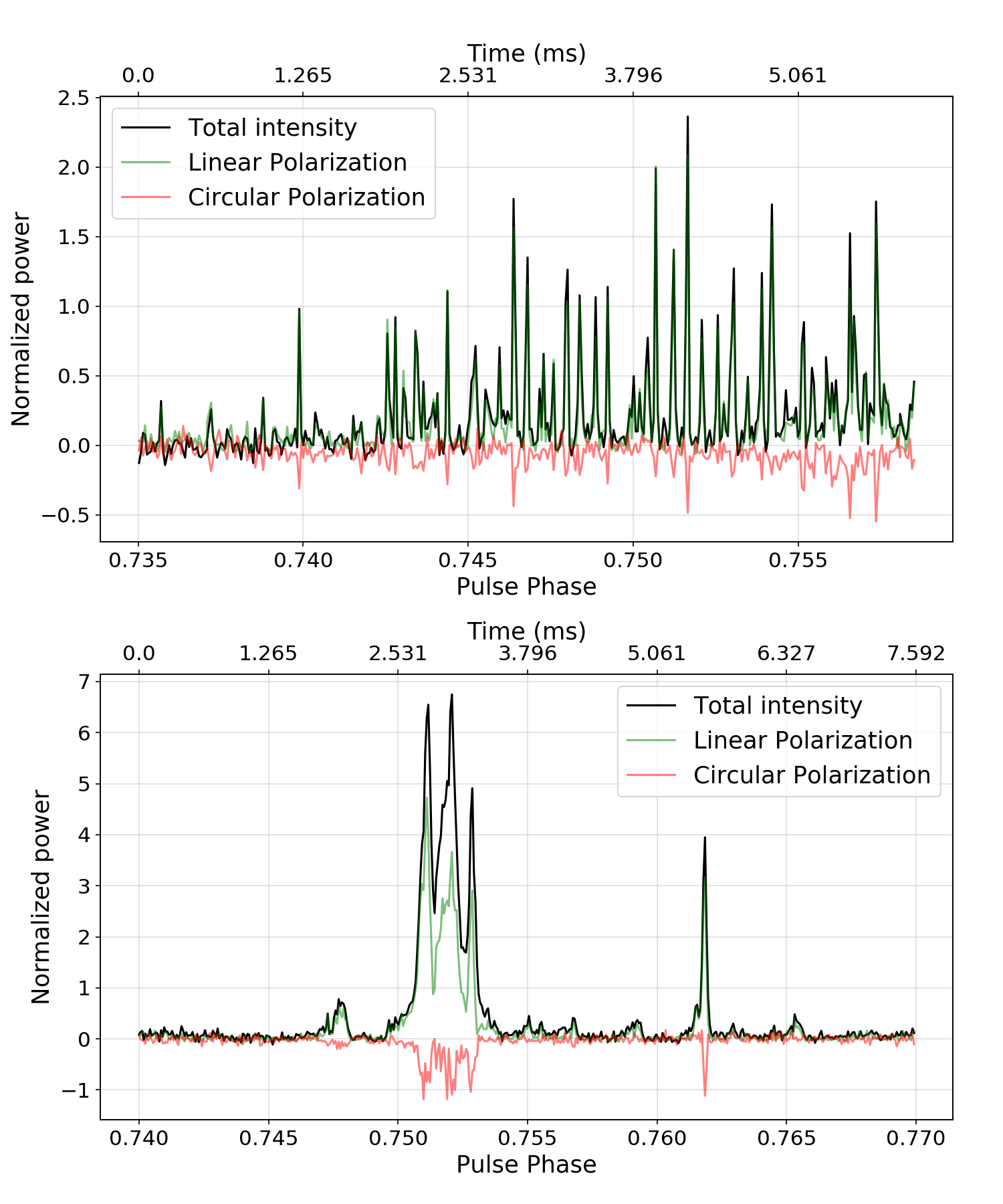}
    \caption{Plot showing `intrinsic' micropulses in a pulse train (upper panel) and also such micropulses appearing alone (lower panel). Plots are normalized by the peak intensity of the folded profile.}
    \label{fig:fig3}
\end{figure}

Some of the single pulses from pulsar B0950+08 show narrow microstructures, that appear without any subpulse beneath them. These microstructures appear in a micropulse train and also alone in a few cases. We used these microstructures as potential tools to study the intrinsic nature of micropulses, uncorrupted by subpulse. We call these micropulses, free from any significant associated subpulse, `intrinsic micropulses'. Fig. \ref{fig:fig3} shows such intrinsic micropulses appearing alone as well as in a micropulse train. Similar to this plot, the X-axis of each single pulse plot in this paper is labeled with both the absolute phase of the pulse window (lower part) and time relative to the start of the pulse window (upper part). We use individual micropulses to study their properties. The process of collecting individual micropulses is the following: We browse through single pulses and mark the micropulses which do not have visible subpulse emission beneath them. We select the time samples in the phase range containing the micropulse (and a few noise samples around it) and use them to study the micropulse properties. We collect a sample of $\sim$500 micropulses from the polarization data with a time resolution of 15.36 $\mu$s to study the polarization properties of `intrinsic' micropulses. Similarly, a sample of more than 800 micropulses from band 3 and a similar sample from band-4 total intensity, wideband data is used to derive the width statistics.
\subsection{Width statistics of `intrinsic' micropulses}\label{sec:sec4.1}
\begin{figure}
    \centering
    \includegraphics[width=\columnwidth]{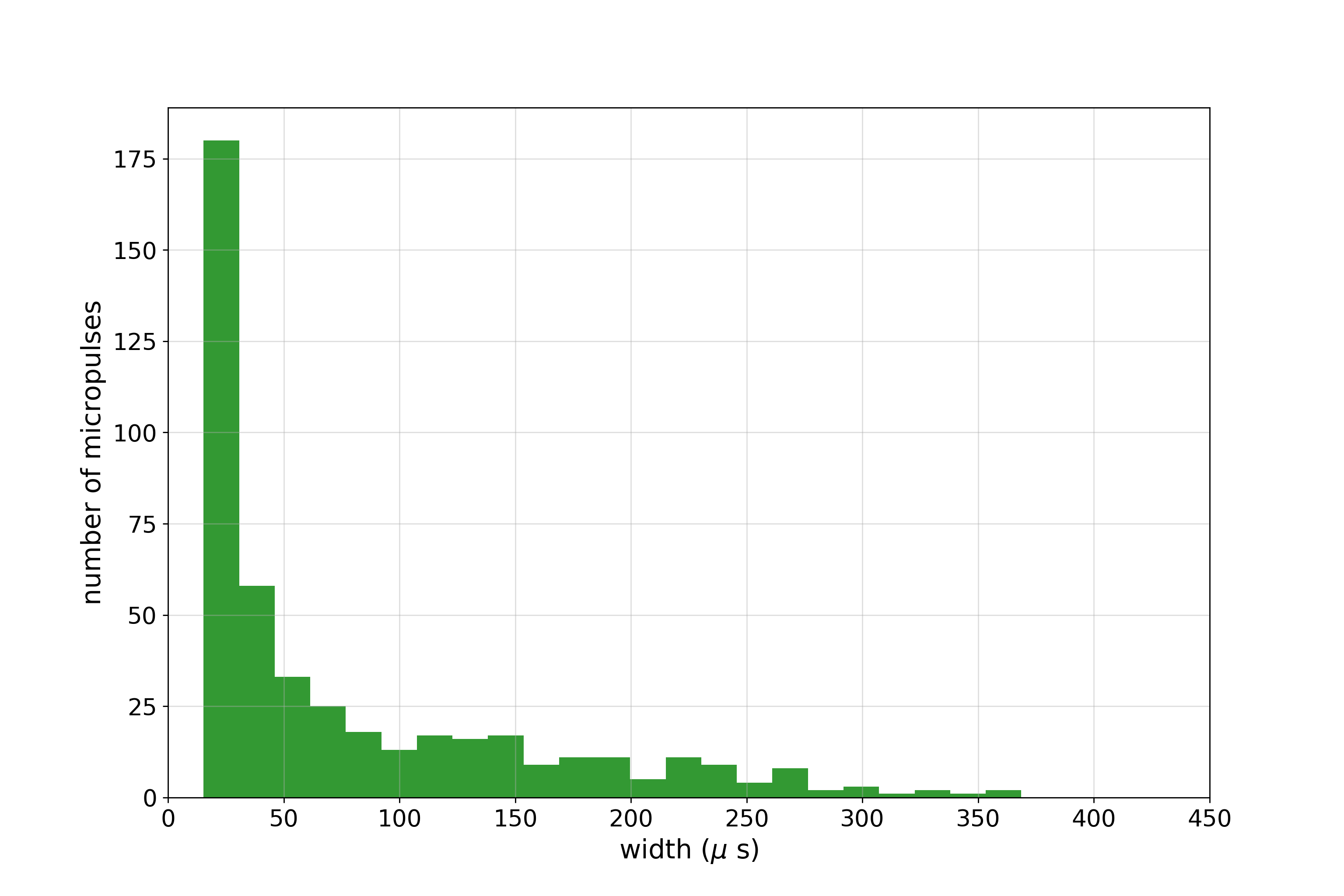}
    \caption{Width histogram of `intrinsic' micropulses selected from polarization data with 15.36 $\mu$s resolution from band-3 (306-338 MHz) of GMRT.}
    \label{fig:fig4}
\end{figure}

We calculate the width of 497 `intrinsic' micropulses collected from polarization data. We use the Stokes-I of the micropulse and calculate the auto-correlation function of the shape of micropulse in total intensity. We use the Full Width at Half Maximum (FWHM) of this auto-correlation function to get the micropulse width. This method is expected to give results similar to the popular method of getting the microstructure width by auto-correlation of micropulse trains \citep{Mitra_2015}. Figure \ref{fig:fig4} shows the histogram of widths for this sample. The bin size is equal to the 15.36 $\mu$s, which is also the time resolution of this sample. The mean width of these micropulses is 81$\pm$15.36 $\mu$s, which is slightly narrower than the micrustructure width ($\sim100$ $\mu$s) reported for this pulsar (\citet{cordes_1990} and \citet{popv_2002}). The distribution peaks at the smallest measurable width (15.36-30.72 $\mu$s) at this resolution and falls rapidly with increasing width. The peak in the histogram tells that the most frequent timescales of micropulses are still unresolved and we need to go to higher time resolutions to get the true timescales of the most frequent micropulses.\\

\begin{figure}
    \centering
    \includegraphics[width=\columnwidth]{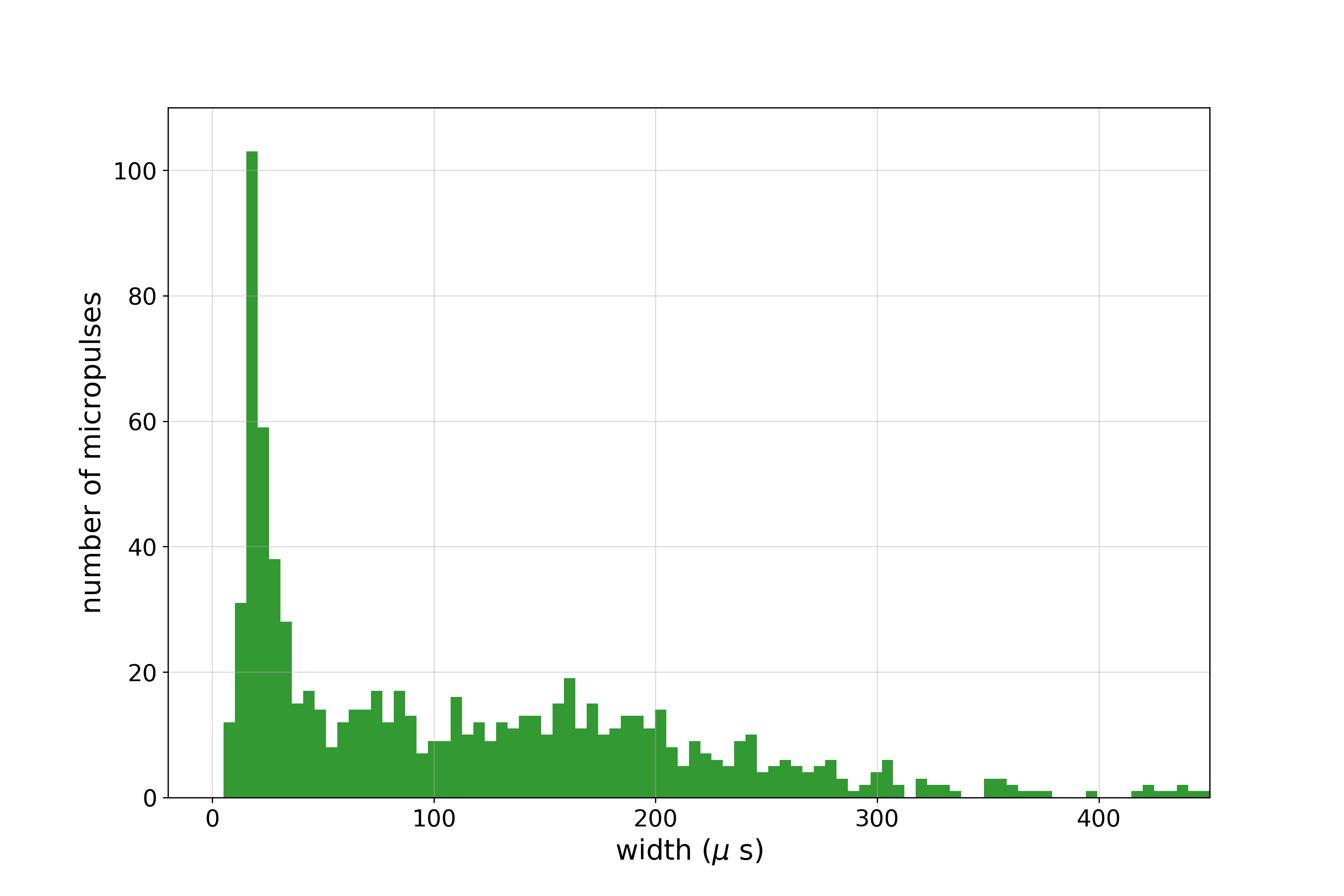}
    \caption{Width histogram of 834 `intrinsic' micropulses from band-3 data (400-500 MHz). The bin size in the histogram is 5.12$\mu$s, the same as the time resolution of wideband data. The histogram peaks at timescales of 15-20 $\mu$s. The average width is 115 $\mu$s.}
    \label{fig:fig5}
\end{figure}

\begin{figure}
    \centering
    \includegraphics[width=\columnwidth]{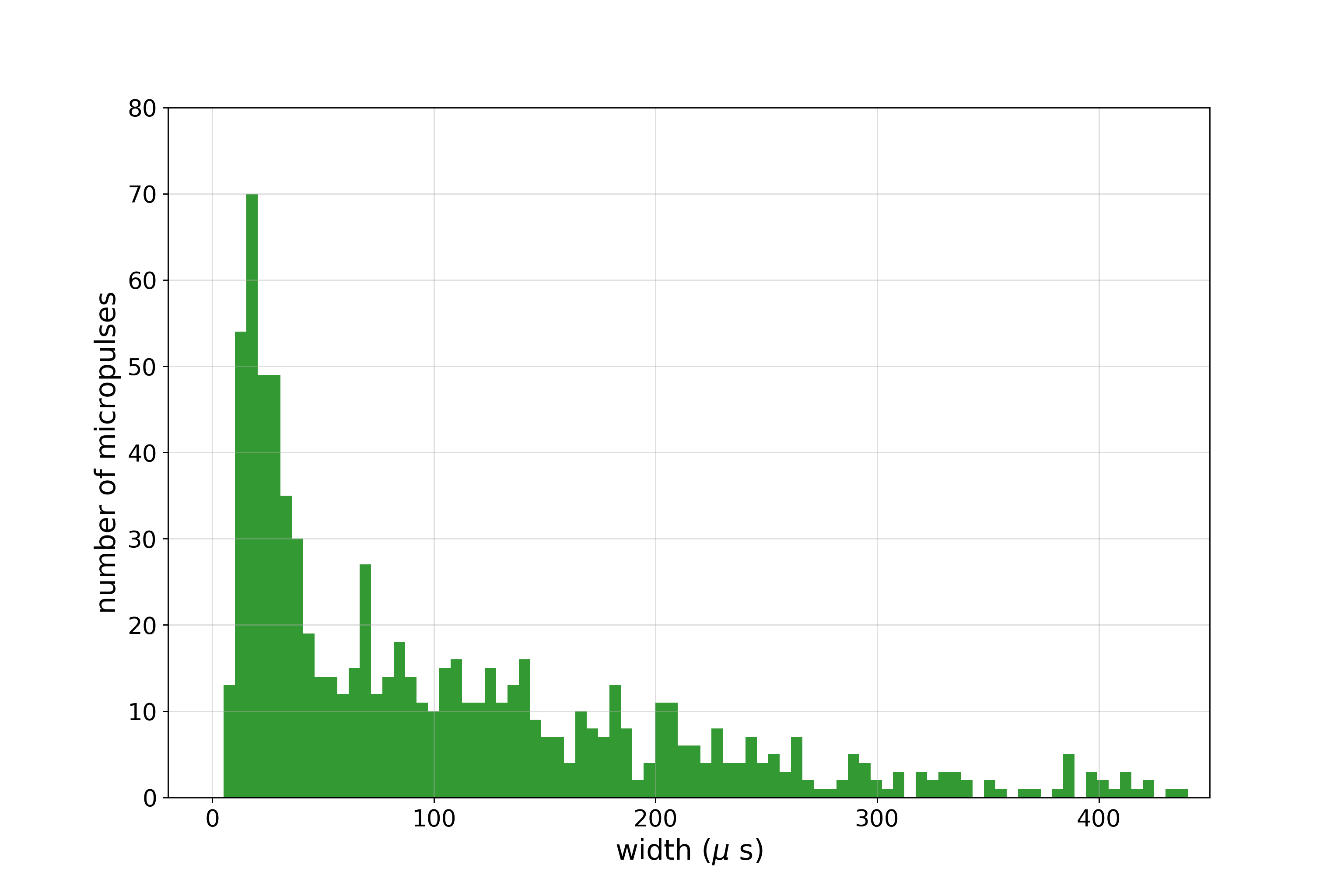}
    \caption{Width histogram of 843 `intrinsic' micropulses from band-4 data (650-750 MHz). The bin size in the histogram is 5.12$\mu$s, the same as the time resolution of wideband data. The histogram peaks at timescales of 15-20 $\mu$s. The average width is 125 $\mu$s.}
    \label{fig:fig6}
\end{figure}

We used wideband observations and the raw voltage data recorded at the Nyquist rate in the background of these wideband observations to achieve time resolutions higher than 15.36 $\mu$s. To compute the width statistics of `intrinsic' micropulses, we collected $\sim$850 `intrinsic' micropulses each from the cleaner parts (upper 100 MHz) of the band-3 (400-500 MHz) and band-4 (650-750 MHz) of the uGMRT wideband data. This wideband data was coherently dedispersed and has a time resolution of 5.12 $\mu$s. Fig. \ref{fig:fig5} shows the width histogram of `intrinsic' micropulses from the band-3 data, while Fig. \ref{fig:fig6} is the width histogram from band-4 data. The population of micropulses from the two frequency bands are of similar nature and peak sharply at 20 $\mu$s time scales and fall rapidly at smaller widths, while extending to 400-500 $\mu$s on the other side. The average width is $115\pm 5.12$ $\mu$s in band-3 and $125\pm5.12$ $\mu s$ in band-4, which is consistent with earlier studies of the general population of microstructures from this pulsar (\citet{cordes_1990} and \citet{popv_2002}). The average width values obtained from the total intensity wideband data are higher than the average micropulse width obtained from the polarization data. The width histograms from the wideband total intensity data also show a relatively larger number of wider micropulses than the width histogram from the polarization data. This difference can be caused by the difference in the S/N and the time resolution of the two data sets or some bias while selecting the `intrinsic' micropulses from the two data sets. Nevertheless, the shape of histograms and values of average micropulse widths from polarization data and total intensity data (both band-3 (400-500 MHz) and band-4 (650-750 MHz)) are broadly consistent with each other. This clearly shows that these narrow structures are indeed wide-band micropulses having a similar width to the general micropulse population from this pulsar. \\

The average micropulse width of around 100 $\mu$s for this pulsar agrees with the known relation between microstructure timescales and pulse period (\citet{Cordes_79} and \citet{kramer_2002}). The histogram of micropulse width shows that the most frequent micropulse time scale is $\sim 20$ $\mu$s, much shorter than the average value. Some earlier studies have also shown that the most frequent microstructure timescales are shorter than the average value (\citet{Mitra_2015} and \citet{popv_2002}) at least in this pulsar. \\

\begin{figure}
    \centering
    \includegraphics[width=\columnwidth]{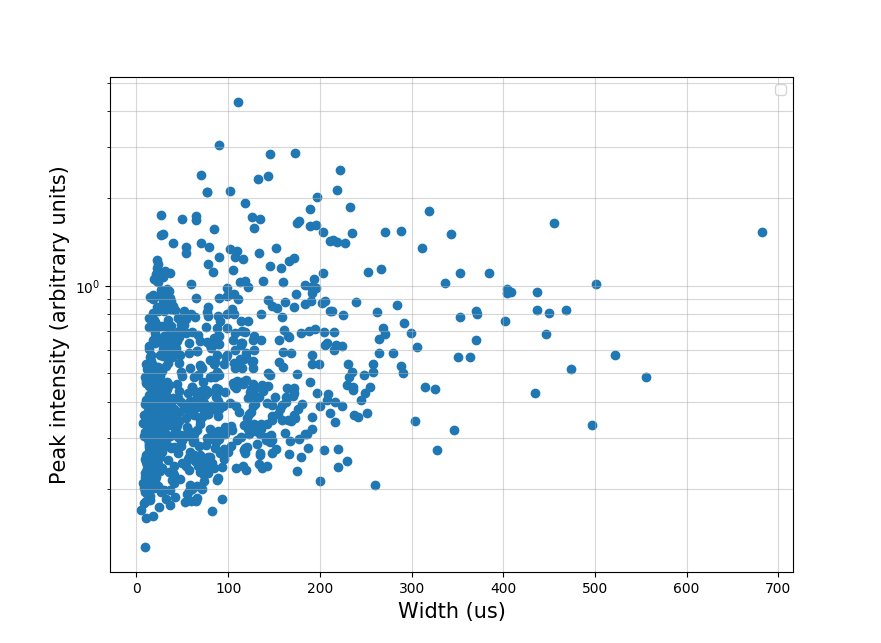}
    \caption{Distribution of width and brightness of 843 `intrinsic' micropulses from band-4 (650-750 MHz) data. The majority of the population is made of faint and narrow micropulses.}
    \label{fig:fig7}
\end{figure}

\begin{figure}
    \centering
    \includegraphics[width=\columnwidth]{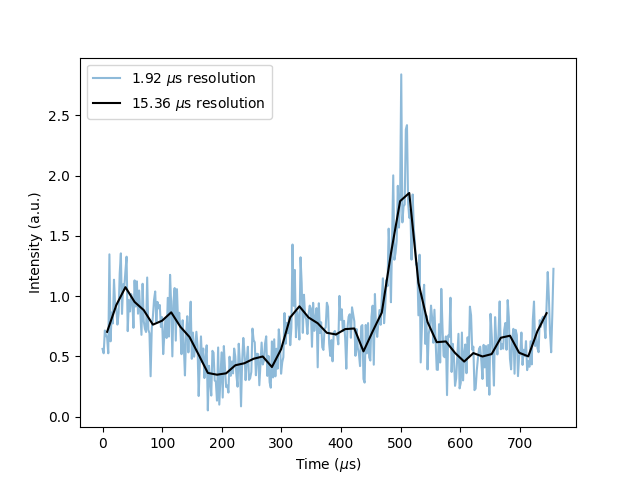}
    \caption{Example of a structure having intrinsic width of 60 $\mu$s. The structure is not breaking into significant multiple narrower components at higher time resolution, indicating that the structure really has a timescale of 60 $\mu$s.}
    \label{fig:fig9}
\end{figure}

The conventional methods of micropulse width calculation are based on calculating the auto-correlation function of single pulses \citep{hankin_nature} or subpulse subtracted single pulses (\citet{Mitra_2015}) to get the timescales of the micropulses. Such methods provide a single value of microstructure timescale for each single pulse, ignoring micropulse width variations within the single pulse. These timescales are naturally dominated by the timescales of brighter micropulses in that particular single pulse. The peak intensity and width distribution of micropulses in Fig. \ref{fig:fig7} shows that majority of micropulses are faint and narrow. These micropulses would get ignored in observations with coarser time resolutions. \cite{kramer_2002} also reported similar dependence of micropulse strength and width for the Vela pulsar. The wider micropulses with very small heights can be easily missed in comparison to their narrower counterparts. This bias can explain the intensity dependence of micropulse width as many micropulses with small intensities and large widths will be missed. We also searched for narrower structures in the high time resolution (0.48 $\mu$s) narrow band data. We could not find any significant structure narrower than 5 $\mu$s. Fig. \ref{fig:fig9} shows a 60 $\mu$s wide micropulse in two time resolutions, 15 $\mu$s and 1.92$\mu$s. The structure does not break into finer structures showing that its intrinsic width really is 60 $\mu$s. In summary, we find that `intrinsic' micropulses have an average width of $\sim$100$\mu$s, while the width distribution peaks at a width of 20 $\mu$s. In our search of the narrowest structures, we did not find any structure having a timescale less than 5$\mu$s.


\begin{figure}
    \centering
    \includegraphics[width=\columnwidth]{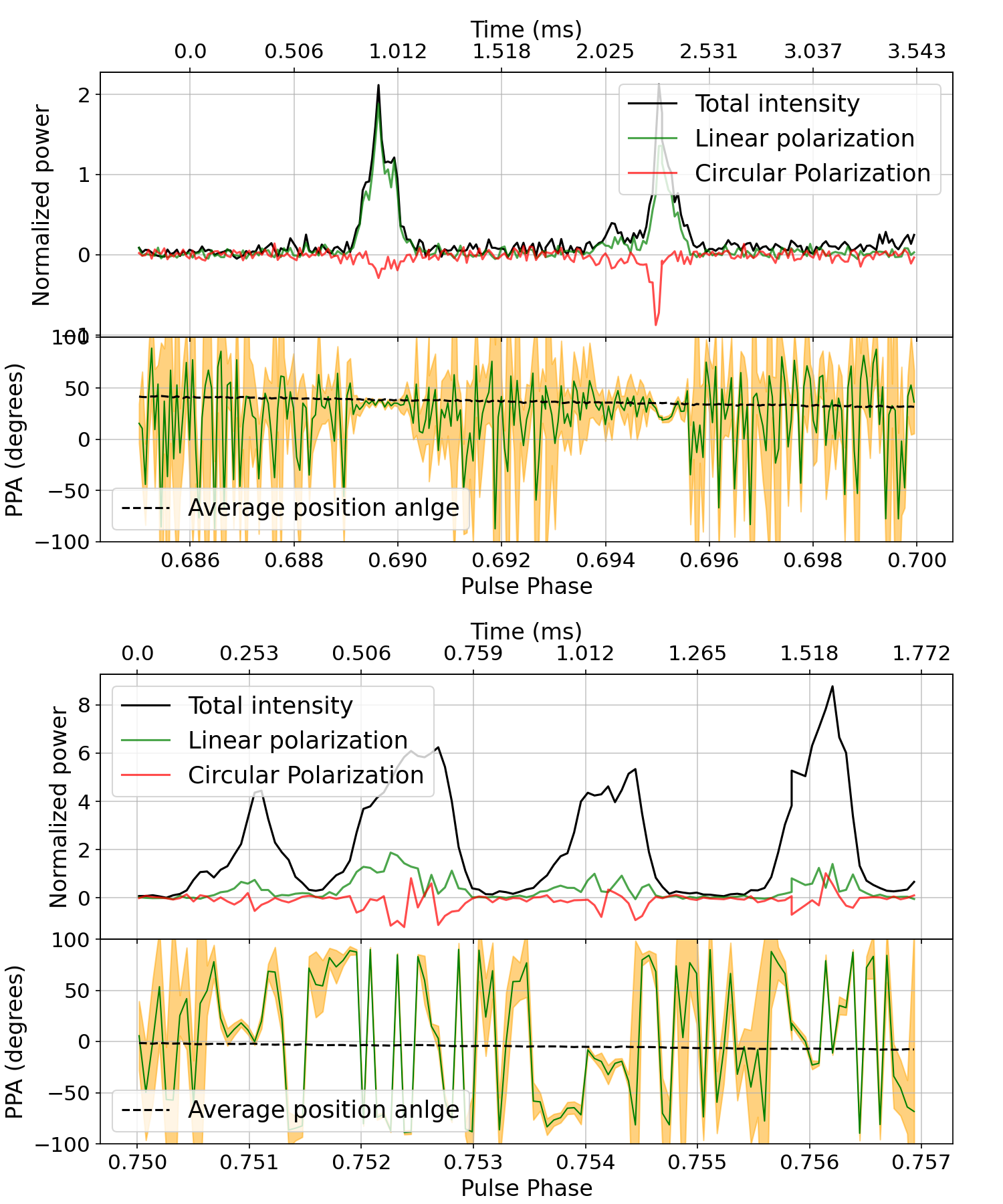}
    \caption{Figure showing `intrinsic' micropulses following the general trend described in Section \ref{sec:sec4.2} (upper panel of the plot), and exceptions to this general trend (lower panel). Orange-shaded regions in the PPA plots represent errors in the PPA. The black dashed curves in the PPA plots are the PPA curve of the folded profile. Plots are normalized by the peak intensity of the folded profile.}
    \label{fig:fig10}
\end{figure}

\begin{figure}
    \centering
    \includegraphics[width=\columnwidth]{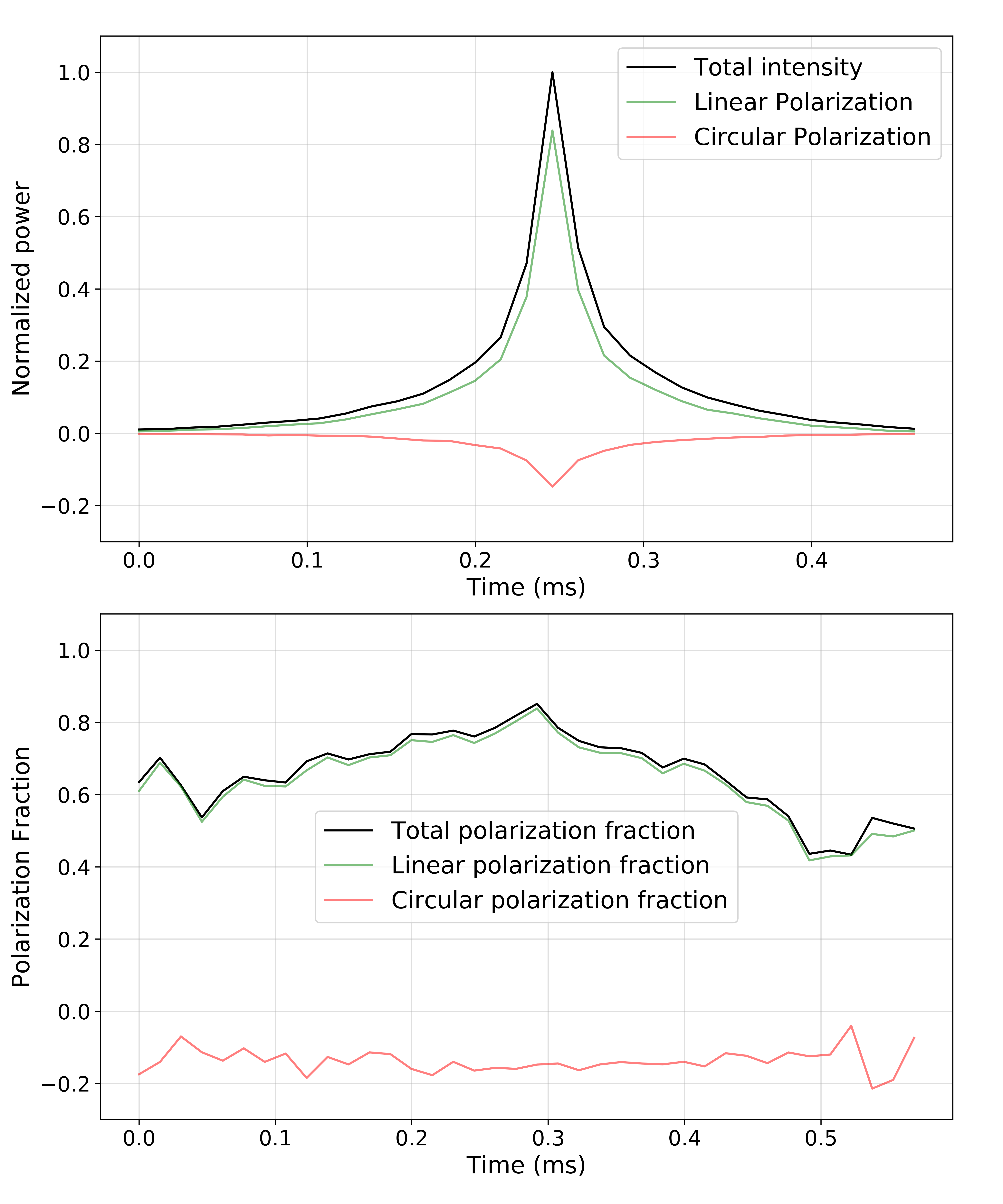}
    \caption{Average shape of 457 `intrinsic' micropulses. The upper panel shows the shape of the profile, while the lower panel shows the polarization fraction across the width of micropulse. }
    \label{fig:fig11}
\end{figure}
\subsection{polarization properties of `intrinsic micropulses'}\label{sec:sec4.2}
Most of the `intrinsic' micropulses are highly polarized. The polarization properties of most of the `intrinsic' micropulses are very similar and this general trend can be described as follows:
\begin{itemize}
    \item All of these `intrinsic' micropulses are highly linearly polarized.
    \item Circular polarization of all these micropulses always has the same sign throughout the micropulse duration (negative in our convention).
    \item These micropulses locally follow the same position angle track given by the folded profile. There are no orthogonal jumps inside a micropulse and all `intrinsic' micropulses are in a single polarization position angle mode. (See upper panel of Fig. \ref{fig:fig10})
\end{itemize}
These polarization properties of `intrinsic' micropulses are partially present in the literature for the general population of micropulses in different pulsars. \citet{cordes_hankin_77}, in their study of four pulsars (including pulsar B0950+08), concluded that the linear polarization of micropulses is stronger than the subpulse and the folded profile. They also showed that microstructure time scales in Stokes I and Stokes V are similar, indicating that there is no sign reversal of circular polarization (circular polarization has the same sign throughout the micropulse duration), which was later confirmed by \citet{Mitra_2015} for a larger sample of pulsars. \citet{cordes_hankin_77} also find that polarization remains roughly constant over the duration of subpulses and micropulses in general. \citet{kramer_2002} showed that there is no sign reversal of circular polarization within a micropulse in the Vela pulsar.\\

In the sample of 497 `intrinsic' micropulses, only $\sim$40 micropulses deviate from this trend (for example, see lower panel of Fig. \ref{fig:fig10}). The major cause of the deviation is depolarization in the wider `intrinsic' micropulses. Hence, it can be safely stated that 90\% of `intrinsic' micropulses follow the above-stated general trend in their polarization properties. It is remarkable that 'intrinsic' micropulses always occur in only one particular mode of the two orthogonal modes, while subpulses can be in either of the two. Given this stable polarization behavior, we prepared an average shape of `intrinsic' micropulses, using 457 micropulses following the trend.
The average shape was obtained by averaging the individual micropulses after aligning their maxima to a single bin. The final average shape is plotted in the upper panel of Fig. \ref{fig:fig11}, while polarization fractions are plotted in the lower panel. Linear polarization fraction remains between 0.6 to 0.8, while circular polarization fraction is remarkably flat at a value of $\sim$0.16 across the profile width. The total polarization fraction reaches 85\% at the peak of the average micropulse shape. \\

Micropulses are often associated with curvature radiation from a relativistically moving point source (single particle or a charge bunch radiating as point source) along the open magnetic field lines. The radiation beam of such point source has an angular width of $\frac{1}{\gamma}$, where $\gamma$ is the Lorentz factor of the relativistically moving point source \citep{lange_1998}. The vacuum curvature radiation from a point source has a characteristic polarization property which includes sign-reversing circular polarization (\citet{Tong_FRB} and \citet{gill_vc}). \citet{Mitra_2015} have shown that the microstructures do not show sign-reversing circular polarization and hence vacuum curvature radiation from a point source does not seem to be the favorable mechanism of micropulse emission. Our study of polarization properties of cleaner individual `intrinsic' micropulses, showing the same sign of circular polarization throughout the micropulse window, also confirms this finding.

\section{Microstructures from pulsar B1642-03}\label{B1642_micro}
\begin{figure}
    \centering
    \includegraphics[width=\columnwidth]{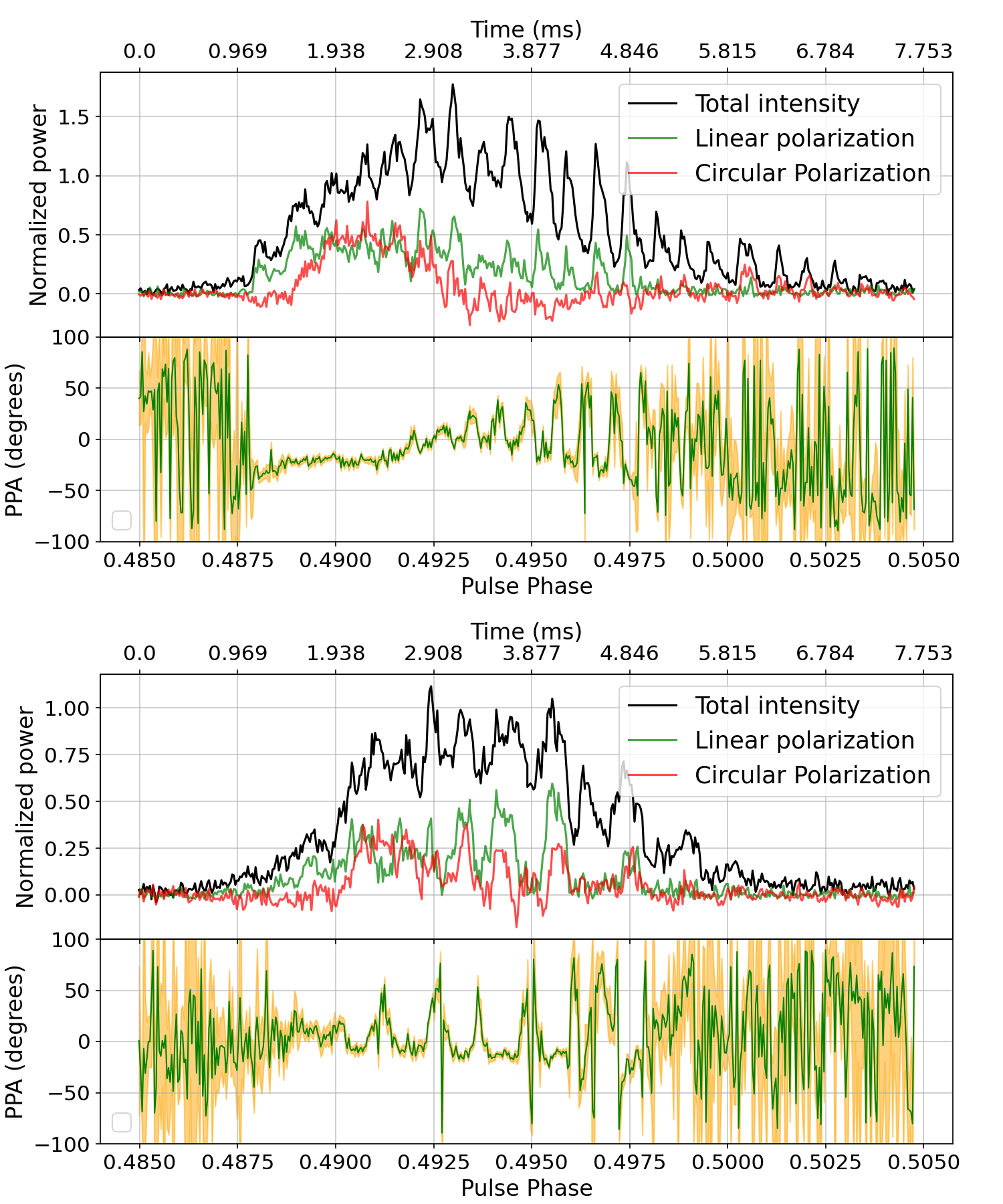}
    \caption{Microstructures in single pulses from pulsar B1642-03. The measured average timescales of the microstructures are shorter in polarization than that in total intensity due to depolarization. Position angle variations caused by micropulses also can be seen. The upper panel shows systematic non-orthogonal jumps caused by the micropulses and the amount of jump being proportional to the strength of micropulse with respect to the subpulse. The lower panel shows another case of micropulse appearing on the subpulse, but this time PPA transitions are close to orthogonal and the linear fraction goes to zero at the phase of transition. The orange-shaded region in the PPA plots shows error in PPA measurements. Plots are normalized by the peak intensity of the folded profile.}
    \label{fig:fig16}
\end{figure}

Now, we move to microstrcures of pulsar B1642-03. pulsar B1642-03 does not show micropulses without subpulses, but it shows quasi-periodic modulations on top of the subpulse emission in some of the single pulses (see Fig. \ref{fig:fig16}). We do not find any study of microstrucutres from this pulsar in the literature. We use the method described by \citet{Mitra_2015} to separate microstructures from the subpulse and calculate the timescale of microstructures. We use a running median filter to get the smooth subpulse. In this method, we use a running box of size 1 ms, shift it by one sample and calculate the median of the box in each step. Once the whole array is covered, the array of median values gives the smooth shape of the subpulse. We subtract this smooth subpulse from the single pulse and the resulting array has microstructures in the form of ripples. Now, we compute the autocorrelation of this resultant array and use the lag corresponding to the first minima in the autocorrelation function as the timescale of the microstructures (see Fig. \ref{fig:micro_procedure}). \\

\begin{figure}
    \centering
    \includegraphics[width=\columnwidth]{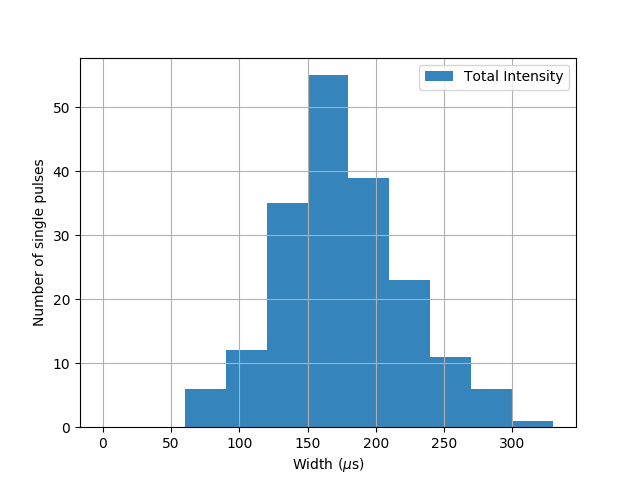}
    \caption{Histogram of microstructure timescales in individual single pulses from pulsar B1642-03. The average microstructure timescale is 180 $\mu$s for this pulsar.}
    \label{fig:micro_1642}
\end{figure}

\begin{figure}
    \centering
    \includegraphics[width=\columnwidth]{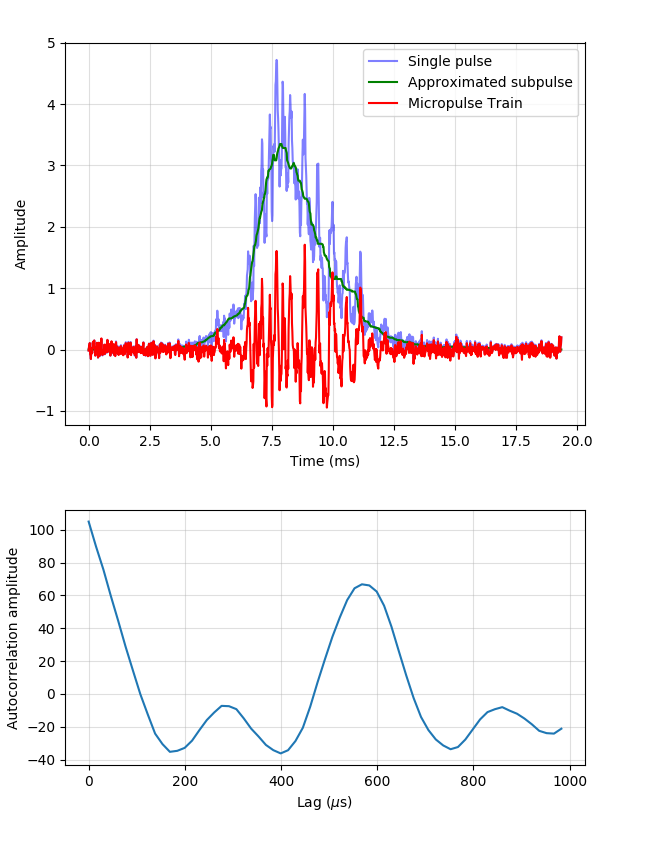}
    \caption{Procedure of measuring timescale of the microstructures from a single pulse. The single pulse, approximated subpulse shape using the running median window, and the micropulse train obtained after subtracting the approximated subpulse from the single pulse are shown in the upper panel. The lower panel shows the autocorrelation function of the micropulse train. The micropulse timescale for this single pulse is 170 $\mu$s.}
    \label{fig:micro_procedure}
\end{figure}

We select 200 bright pulses with clear quasiperiodic modulations on top of subpulses. We have the total intensity, linear polarization fraction, and circular polarization fraction of these single pulses. We compute the timescales of the microstructures as described above in total intensity, linear polarization, and circular polarization. We get the average timescale of microstrucutres to be $180\pm 15 \mu$s in total intensity, $165\pm 15\mu$s in linear polarization, and $137\pm 15 \mu$s in circular polarization. The microstructure timescales are shorter in the linear and circular polarization due to the high depolarization in the single pulses (as seen in Fig. \ref{fig:fig16}). Fig. \ref{fig:micro_1642} shows the histogram of the microstructure total intensity timescales in the single pulses from this pulsar. The peak of the histogram is 150 $\mu$s, which is similar to the average value of the microstructure timescale in this pulsar. The micropulse width of $180 ~\mu s$ in this pulsar is consistent with the relation between micropulse width and the period of the pulsar. \\

Two kinds of timescales for micrstrucutres are reported in the literature. The first one is a measurement of the width of the microstructures, which either corresponds to the break in the autocorrelation function of the single pulse \citep{hankin_nature} or the first minimum in the autocorrelation function of the subpulse subtracted single pulse \citep{Mitra_2015}. In this work, we have reported the width of individual `intrinsic' micropulses from the pulsar B0950+08, which has been measured as the FWHM of the autocorrelation function of the individual micropulses. The FWHM of the autocorrelation function is comparable to the width at $10\%$ height of the original micropulse, if the micropulse has a Gaussian shape. The second timescales reported for the microstructures is the periodicity of the micropulses in the micropulse train. This periodicity corresponds to the lag of the first peak in the auto-correlation function \citep{Mitra_2015}.   

\section{Single pulse polarization properties of pulsars B0950+08 and B1642-03}\label{sec:sec5}

Now, we will try to understand the polarization behavior of single pulses from pulsars B0950+08 and B1642-03 using some simple empirical models. In section \ref{sec:sec4.2}, We saw that `intrinsic' micropulses have very stable polarization properties and 90\% of these micropulses belong to a particular orthogonal mode, which is also seen in the folded profile and hence should be the dominant one of the two possible orthogonal modes. \citet{cordes_hankin_77} found that the polarization properties are roughly stable within a subpulse or micropulse and orthogonal jumps are often seen at the edges of micropulses or subpulses. Single pulses from pulsars B0950+08 and B1642-03 show a diverse range of polarization properties. The linear and circular polarization fractions are highly variable and can vary even within a subpulse. Position angle transitions are often seen at the junctions of two subpulses or junctions of micropulses and subpulses as reported by \citet{cordes_hankin_77}.  These transitions can be either full 90$^{\circ}$ jumps or non-orthogonal jumps with less than 90$^{\circ}$ change in position angle. The micropulses on the top of subpulses make very interesting cases of mode mixing. The position angle jumps occurring at the junctions of subpulse-subpulse and subpulse-micropulse cases, indicate that two mixing components (two subpulses or subpulse and micropulses on the top of it) are in different position angle modes. When the linear polarization of one component starts dominating the other component, position angle jumps happen. \\

The nature of position angle transitions (orthogonal/non-orthogonal) is different for the two pulsars in our sample. While pulsar B0950+08 almost always shows orthogonal position angle transition at the junctions of two components, pulsar B1642-03 usually shows both orthogonal and non-orthogonal jumps at such locations. Smooth variations in position angle are sometimes accompanied by similar variations in the circular polarization fractions in both pulsars (see lower panel of figure \ref{fig:fig14} and upper panel of \ref{fig:fig15}). \\

\begin{figure}
    \centering
    \includegraphics[width=\columnwidth]{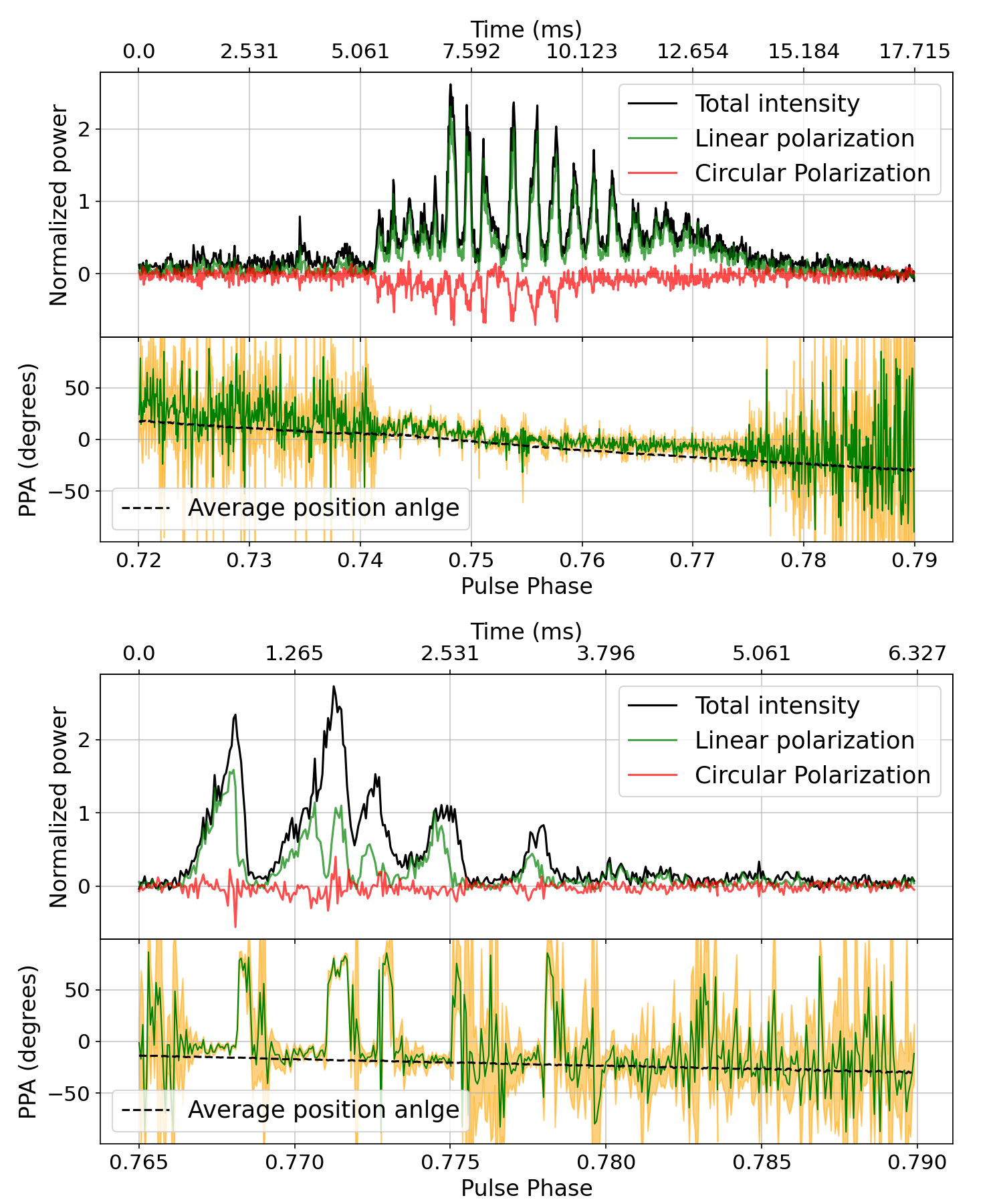}
    \caption{Micropulses appearing with subpulses in pulsar B0950+08. The upper panel shows a case of fully polarized subpulse along with fully polarized micropulses riding on top of it. The lower panel shows overlapping micropulses resulting in depolarization and orthogonal jumps within the micropulses. The orange-shaded regions in the PPA plots represent error in PPA estimates. Plots are normalized by the peak intensity of the folded profile.}
    \label{fig:fig13}
\end{figure}

\begin{figure}
    \centering
    \includegraphics[width=\columnwidth]{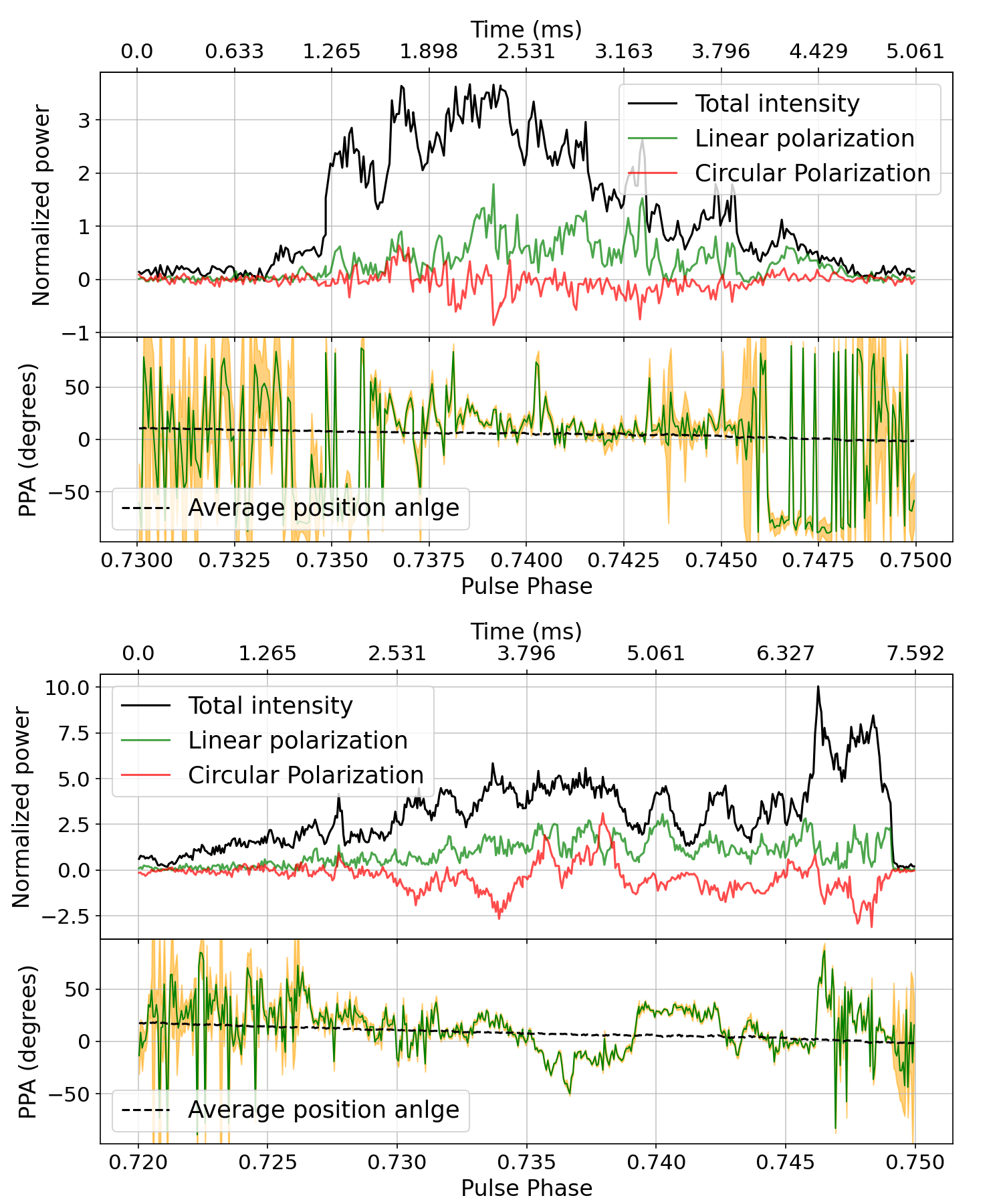}
    \caption{Position angle transition in subpulses from pulsar B0950+08. The upper panel shows an orthogonal jump within the subpulse which can be very well explained by the incoherent superposition of two orthogonal modes. The lower panel shows clear non-orthogonal position angle transitions within the subpulse, these non-orthogonal transitions can not be explained by the incoherent superposition of two orthogonal modes. The orange-shaded regions in the PPA plots represent errors in PPA estimates. Plots are normalized by the peak intensity of the folded profile.}
    \label{fig:fig14}
\end{figure}

Let us consider PPA transitions happening at the junctions of micropulses and the subpulses at which the micropulse is riding. In some of these cases, the mode of subpulse and micropulse happens to be the same, resulting in very high polarization and no transition in PPA (see the upper panel of fig. \ref{fig:fig13}). But if the modes of micropulse and subpulse are different then, either orthogonal (see lower panel of fig. \ref{fig:fig13} and \ref{fig:fig16}) or non-orthogonal (upper panel of fig. \ref{fig:fig16}) jumps are seen at the edges of micropulses. The amount of PPA jumps in the cases of non-orthogonal jumps seems to be proportional to the ratio of the linear fraction of two components at the junction (see the upper panel of fig. \ref{fig:fig16}). Often significant depolarization of subpulse is seen in such cases. The linear polarization fraction goes to zero at each sharp orthogonal jump (lower panels of fig. \ref{fig:fig13}, \ref{fig:fig16} and the upper panel of fig. \ref{fig:fig14}).\\

Figure \ref{fig:fig15} shows two cases of PPA transitions from pulsar B1642-03, caused by the overlap of two components of subpulses with different modes. The lower panel shows a subpulse that has a weak component overlapping with a strong component. The weak component suddenly gets dominated by the strong one and we see a sudden sharp change in the position angle at the junction of these two components. The upper panel shows two components of subpulse with comparable strengths but have different position angle modes and sense of circular polarization. The junction point of these two components can be identified by the sharp increase in intensity when the second component starts. Since power in linear polarization is similar in both components, the second component gradually starts dominating the first one, and we see a slow and gradual transition in position angle. The upper panel of Fig. \ref{fig:fig14} shows a case of mixing of subpulses in different PPA modes from pulsar B0950+08. The linear polarization clearly goes to zero at the locations of orthogonal jumps.\\

Two waves with different modes of polarization can combine in two ways, the first is incoherent superposition when waves don't have any phase relation, and the second is coherent superposition when there is a definite phase relation between the two waves. We will look at these two possible scenarios and apply them to understand the observed polarization properties of single pulses. 
\subsection{Incoherent superposition of two polarized waves}
\begin{figure}
    \centering
    \includegraphics[width=\columnwidth]{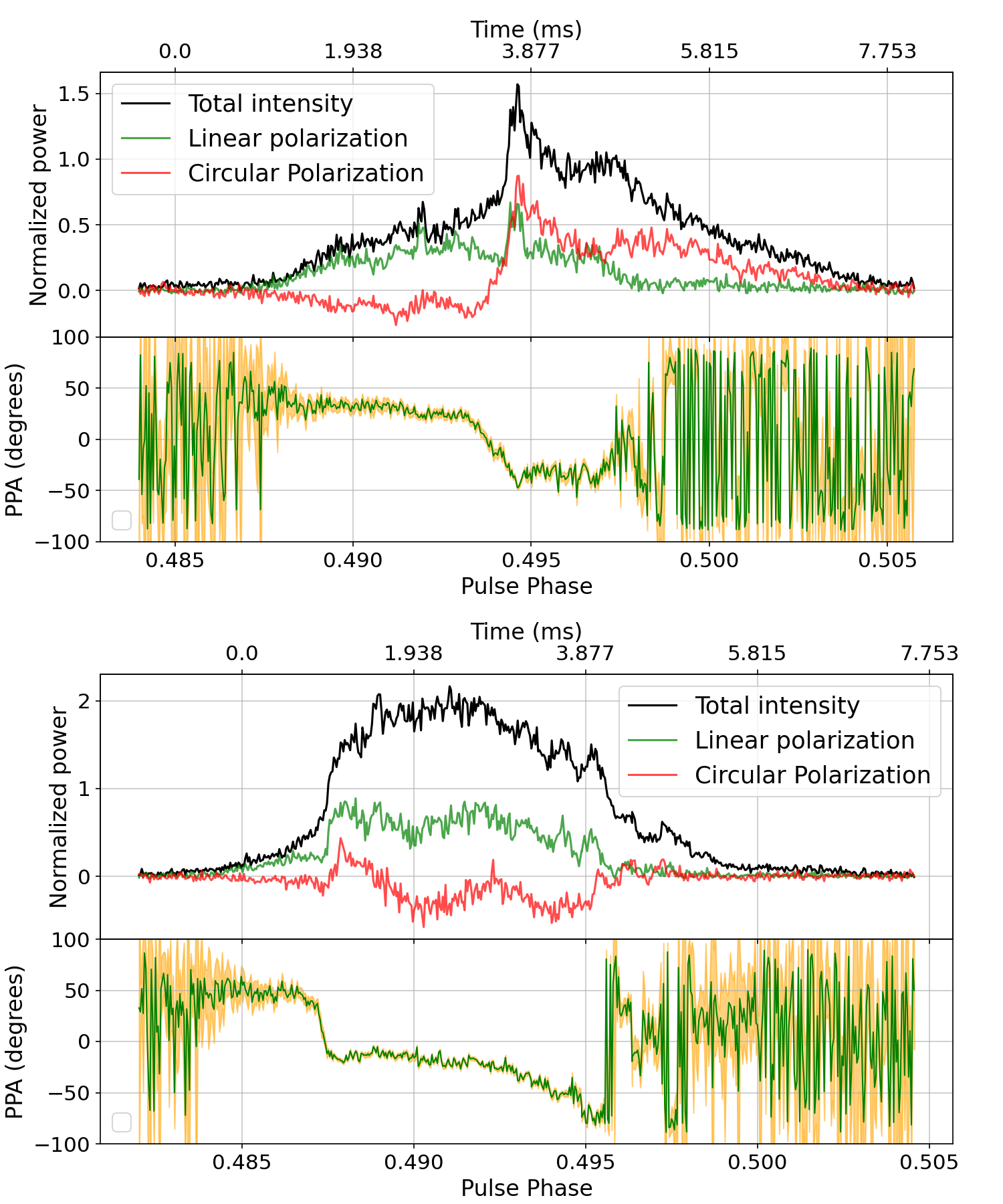}
    \caption{Position angle transition in subpulses of pulsar B1642-03. The orange-shaded regions in the PPA plots represent errors in PPA estimates. The upper panel shows a smooth change in polarization properties and transition in PPA. The lower panel shows a sharp transition in total intensity, linear polarization, circular polarization, and position angle. But the transition is not orthogonal and also linear polarization fraction is not going to zero at the phase of PPA transition, it can not be explained by incoherent superposition. A coherent superposition of two orthogonal modes is required to explain such PPA transitions. Plots are normalized by the peak intensity of the folded profile.} 
    \label{fig:fig15}
\end{figure}

In this model, we consider two orthogonal modes and their relative strength to vary with the rotation phase. It is seen often that different components of a single pulse have different PPA mode. Also, we saw that 'intrinsic' micropulses from pulsar B0950+08 mostly follow a unique PPA traverse. So, here we hypothesize that subpulses are made up of one or more than one component that can have different modes of PPA, and orthogonal jumps happen when two of such components having orthogonal PPA modes overlap. Sometimes the presence of two or more subpulse components is clearly visible in the intensity gradients (see Figure \ref{fig:fig15}). Micropulses also can have different PPA mode than the subpulses on which they are riding, this generates microstructure timescale variations in the PPA traverse of subpulse consistent with the locations of micropulses (see fig. \ref{fig:fig16}). Such `wiggles' of microstructure timescales were also reported by \citet{kramer_2002} for the Vela pulsar. Here we will see what happens when two components with different PPA modes overlap and there is no phase relation between the two components.\\

\begin{figure}
    \centering
    \includegraphics[width=\columnwidth]{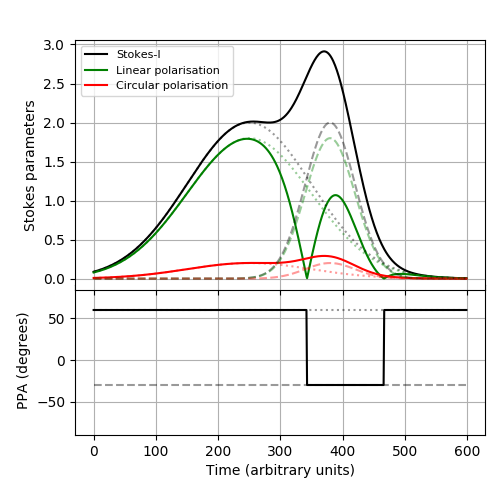}
    \caption{Illustration of incoherent superposition of two subpulse components.}
    \label{fig:fig12}
\end{figure}

If two elliptically polarized waves with orthogonal polarization angles are superposed with no phase relation between them, then Stokes parameters of the resulting wave are given as follows (see appendix for detailed derivation),
$$I=I_1+I_2$$
$$Q=Q_1+Q_2$$
$$U=U_1+U_2$$
$$V=V_1+V_2$$
$$\psi=\psi_1~~ or ~~\psi_2$$
Where $(I_1,Q_1,U_1,V_1)$ and $(I_2,Q_2,U_2,V_2)$ are Stokes parameters of two superposing waves. $\psi_1$ and $\psi_2$ are the position angles of these two waves, where $\psi_2=\psi_1+\pi/2$. For these two orthogonal waves $Q_1$, $Q_2$, and $U_1$, $U_2$ will have opposite signs since the orthogonal position angle will introduce 180 degrees rotation in the Q-U plane. Fig. \ref{fig:fig12} illustrates the incoherent superposition of two components of a subpulse. These components are shown by dotted and dashed curves. Initially, the first mode (represented by dotted curves) dominates, but slowly second mode (represented by dashed curves) starts to rise. Since the U and Q of these two modes have opposite signs, they begin to cancel out each other. When the linear polarization contribution of two modes is equal, the U and Q of the resultant go to zero. This can be understood as an incoherent superposition of two orthogonal fully linearly polarized waves with equal amplitude that will create a perfectly unpolarized signal. Till this point, Some fraction of the linear polarization of the first mode was combining incoherently with the linear polarization of the second mode to produce the unpolarized signal, but the remaining part of the first mode was still polarized with a position angle corresponding to the first mode. This explains the depolarization and PPA remaining in the first mode. After this point, the second mode dominates and hence remaining part is having PPA of the second mode. Then at some point, the first mode again starts dominating and PPA goes back to the first mode. So, the incoherent superposition of two modes naturally explains the depolarization and the linear fraction going to zero at the sites of orthogonal PPA jumps.\\

The 'intrinsic micropulses' and their remarkable stable polarization properties were discussed in section \ref{sec:sec4.2}. It should be noted that around $10\%$ of the intrinsic micropulses deviate from the general trend in the polarization properties. The main cause of the deviation is depolarization in wider micropulses. These exceptions to the general trend can be comfortably explained by considering the incoherent superposition of two narrow components having orthogonal PPA modes. Though this superposition model explains orthogonal jumps, depolarization, and linear polarization going to zero just before the orthogonal jump, it fails to provide any explanation for the existence of circular polarization, systematic non-orthogonal jumps in the PPA, and smooth variations in PPA correlated with the circular polarization seen in both pulsars in our sample. 


\subsection{Coherent superposition of two polarized waves}
The coherent superposition of two orthogonal modes is sometimes used to explain the polarization angle jumps and the existence of circular polarization \citep{dykes_2018}. It also becomes necessary to explain the systematic non-orthogonal jumps seen in pulsars (see the upper panel of figure \ref{fig:fig16}). Non-orthogonal jumps are not allowed in incoherent superposition. Coherent superposition also allows the conversion of linear into circular and vice-versa, depending on the phase difference between the two waves, providing means to explain the cases of correlation between circular polarization and PPA variations. But the depolarization can not be explained by the coherent superposition of two orthogonal modes.\\

In our set of two pulsars, pulsar B0950+08 has both highly polarized and depolarized single pulses. The jumps in pulsar B0950+08 are mostly orthogonal in the depolarized regions. These jumps agree with the incoherent superposition model explained in the previous subsection. All the highly linearly polarized ($>85\%$) subpulses only have one PPA mode, which is very close to the PPA traverse of the folded profile (see the upper panel of figure \ref{fig:fig13}). The other mode is often seen in depolarized subpulses. On the other hand, pulsar B1642-03 often shows systematic non-orthogonal jumps associated with the presence of micropulses and other components of the subpulse. This also shows orthogonal jumps which are not associated with linear going to zero before the jumps (See figure \ref{fig:fig15}), which again indicates towards coherent superposition of two modes. In general, the incoherent superposition model is sufficient to explain the depolarization and orthogonal jumps in the pulsar B0950
+08, while coherent superposition is required to explain the systematic non-orthogonal jumps in the pulsar B1642-03 and correlated variations in circular polarization and polarization position angle seen in both pulsars.

\section{Summary}\label{sec:summary}
Microstructures in general appear as quasi-periodic narrow structures on the top of subpulses. Due to the presence of associated subpulse emission, there is an inherent difficulty in measuring the true properties of microstructures, be it timescales, strength, or polarization properties. We studied isolated micropulses in single pulses from pulsar B0950+08 to probe the true nature of micropulse emission. We call these isolated micropulses `intrinsic' micropulses. We collected a sample of $\sim500$ `intrinsic' micropulses with polarization information. We find that $\sim90\%$ of the `intrinsic' micropulse population follows a characteristic polarization property. They all are highly linearly polarized, with the position angle following the polarization angle track of the folded profile. Circular polarization is always negative for all these `intrinsic' micropulses following the characteristic polarization behavior. This finding hints towards a similar emission process for all micropulses. We also measured micropulse timescales from samples of $\sim$850 micropulses in both band 3 (400-500 MHz) and band 4 (650-750 MHz) of uGMRT. We find that the width distribution of micropulses in both frequency bands is similar. We also notice that while the average micropulse width is close to $\sim 100 \mu s$, the most frequently appearing timescale is $\sim 20\mu s$, much smaller than the average value. We also argue that the conventional way of finding characteristic timescales of microstructure gives the average width of micropulses and ignores the variations of the timescales within a given single pulse. We also report the timescales of microstructures from pulsar B1642-03 to be $\sim 180\mu s$. Both these measurements of average microstructure timescales are consistent with the known relation between microstructure width and the period of pulsars.\\

The polarization position angle of the folded profile of many pulsars traverse S-shaped curves. Two orthogonal modes of emissions are usually seen in pulsars \citep{Mitra_2017}. Mixing of these modes is often used to explain the deviations of PPA from the characteristic S-curve in single pulses. To explain the variations, a single pulse is considered to be made of superposition of many components, and each component can be in any of the two orthogonal modes. We find that most of the polarization position angle jumps in pulsar B0950+08 can be explained by considering an incoherent superposition of two orthogonal modes. But coherent superposition is needed to explain more complex and systematic non-orthogonal jumps occasionally seen in pulsar B0950+08 and regularly seen in pulsar B1642-03. Coherent superposition also provides means to convert linear polarization into circular and may help to understand the correlation in variations of polarization position angle and circular polarization.

\section*{Acknowledgements}
 The GMRT is run by the National Centre for Radio Astrophysics of the Tata Institute of Fundamental Research, India. We acknowledge the support of GMRT telescope operators and the GMRT staff for supporting the GHRSS survey observations.
\section*{Data Availability}
The data underlying this article will be shared on reasonable request to the corresponding author.



\bibliographystyle{mnras}
\bibliography{example} 




\appendix
\section{Derivation of incoherent superposition of two orthogonal modes}
\begin{figure}
    \includegraphics[width=\columnwidth]{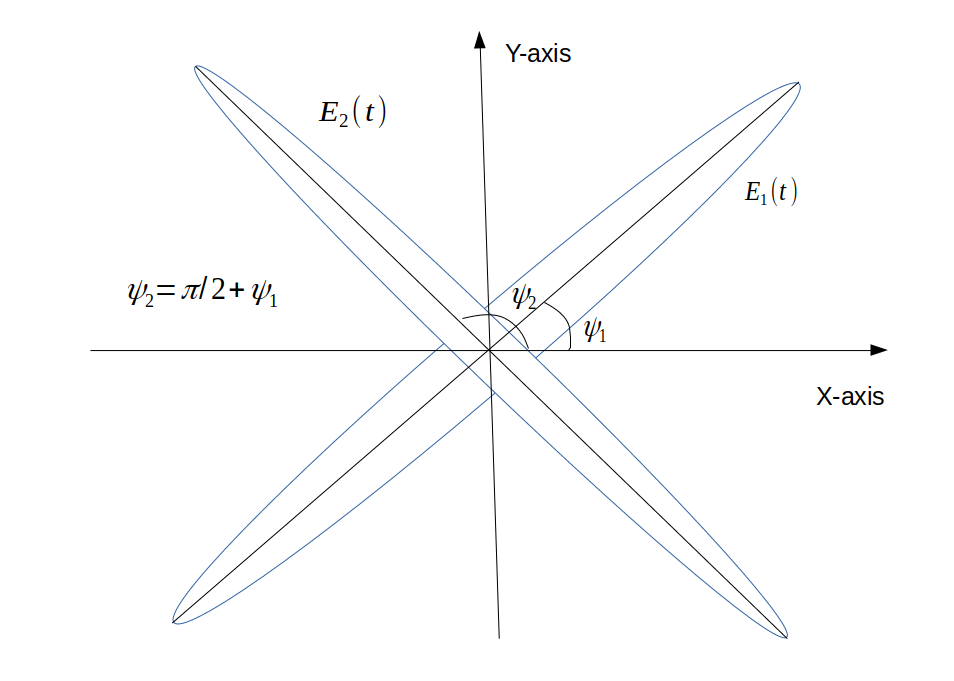}
    \caption{Schematic diagram showing two elliptically polarized orthogonal modes. These two waves have position angles separated by 90$^0$. }
    \label{fig:A1}
\end{figure}

Elliptically polarized waves can be written as a superposition of two orthogonal linearly polarized waves with a phase difference between them. 
$$\vec{E(t)}=e^{i\delta}(E_{0}\cos\alpha\hat{x}+E_{0}\sin\alpha e^{i\beta}\hat{y})e^{i(\omega t-kz)}$$
 Here, $E_{0}\cos\alpha$ and $E_{0}\sin\alpha$ are the amplitudes of the two orthogonal plane waves with a phase difference of $\beta$, that after superposition make the elliptically polarized wave $\vec{E(t)}$.
The Stokes parameters of this wave will be given by,
$$I=<E_xE_x^*>+<E_yE_y^*>=E_0^2$$
 $$Q=<E_xE_x^*>-<E_yE_y^*>=E_0^2\cos 2\alpha$$
 $$U=<E_xE_y^*>+<E_x^*E_y>=E_0^2 \sin 2\alpha\cos\beta$$
 $$V=\frac{1}{i}(<E_xE_y^*>-<E_x^*E_y>)=-E_0^2\sin 2\alpha \sin \beta$$
 And the position angle ($\psi$) will be,
 $$\tan 2\psi=\frac{U}{Q}=\tan 2\alpha \cos\beta$$
In our case, we have two elliptically polarized waves. Position angles of these waves are separated by $\pi/2$ (see figure \ref{fig:A1}). 

 The electric field for the first elliptical wave is given by,
 $$\vec{E_1(t)}=e^{i\delta_1}(E_{10}\cos\alpha_1\hat{x}+E_{10}\sin\alpha_1 e^{i\beta_1}\hat{y})e^{i(\omega t-kz)}$$
 Here, $E_{10}\cos\alpha_1$ and $E_{10}\sin\alpha_1$ are the amplitudes of the two orthogonal plane waves with a phase difference of $\beta_1$, that after superposition make the elliptically polarized wave $\vec{E_1(t)}$. Similarly, a second elliptically polarized wave can be written as,
 $$\vec{E_2(t)}=e^{i\delta_2}(E_{20}\cos\alpha_2\hat{x}+E_{20}\sin\alpha_2 e^{i\beta_2}\hat{y})e^{i(\omega t-kz)}$$
 Here $\delta_1-\delta_2$ denotes the phase difference between $E_1$ and $E_2$ and this must jump randomly between $0-2\pi$ at time scales of $\frac{1}{\omega}$, in order to make these waves incoherent. The phase difference between X and Y components of these elliptical waves are given by $\beta_1$ and $\beta_2$.\\ 
 Stokes parameters of these two waves will be given as,
 $$I_1=E_{10}^2$$
 $$Q_1=E_{10}^2\cos 2\alpha_1$$
 $$U_1=E_{10}^2\sin 2\alpha_1 \cos \beta_1$$
 $$V_1=-E_{10}^2 \sin 2\alpha_1 \sin \beta_1$$
 $$\psi_1=\frac{1}{2}\tan^{-1}(\tan2\alpha_1\cos\beta_1)$$

 $$I_2=E_{20}^2$$
 $$Q_2=E_{20}^2\cos 2\alpha_2$$
 $$U_2=E_{20}^2\sin 2\alpha_2 \cos \beta_2$$
 $$V_2=-E_{20}^2 \sin 2\alpha_2 \sin \beta_2$$
 $$\psi_2=\frac{1}{2}\tan^{-1}(\tan2\alpha_2\cos\beta_2)$$
 And, $\psi_2=\psi_1+\frac{\pi}{2}$.\\
 Now, x and y components of the resultant of these two waves will be given by,
 $$E_x=(e^{i\delta_1}E_{10}\cos\alpha_1+e^{i\delta_2}E_{20}\cos\alpha_2)e^{i(\omega t-kz)}$$
 $$E_y=(e^{i\delta_1}E_{10}e^{i\beta_1}\sin\alpha_1+e^{i\delta_2}e^{i\beta_2}E_{20}\sin\alpha_2)e^{i(\omega t-kz)}$$
 Now, given $E_x$ and $E_y$, We can calculate Stokes parameters.
 $$I=<E_xE_x^*>+<E_yE_y^*>$$
 \begin{math}
 =<E_{10}E_{10}^*>\cos^2\alpha_1+<E_{20}E_{20}^*>\cos^2\alpha_2\\+<E_{10}E_{10}^*>\sin^2\alpha_1+<E_{20}E_{20}^*>\sin^2\alpha_2
\end{math}
\begin{equation}
    \boxed{I=I_1+I_2}
\end{equation}
$$Q=<E_xE_x^*>-<E_yE_y^*>$$
$$=<E_{10}E_{10}^*>(\cos^2\alpha_1-\sin^2\alpha_1)+<E_{20}E_{20}^*>(\cos^2\alpha_2-\sin^2\alpha_2)>$$
\begin{equation}
    \boxed{Q=I_1\cos2\alpha_1+I_2\cos2\alpha_2=Q_1+Q_2}
\end{equation}
 $$U=<E_xE_y^*>+<E_x^*E_y>$$
 \begin{math}
 =I_1\sin\alpha_1\cos\alpha_1e^{-i\beta_1}+I_2\cos\alpha_2\sin\alpha_2e^{-i\beta_2}+I_1\sin\alpha_1\cos\alpha_1e^{i\beta_1}+I_2\cos\alpha_2\sin\alpha_2e^{i\beta_2}
 \end{math}
 
 \begin{equation}
     \boxed{U=I_1\sin2\alpha_1\cos\beta_1+I_2\sin2\alpha_2\cos\beta_2=U_1+U_2}
 \end{equation}
 $$V=\frac{1}{i}(<E_xE_y^*>-<E_x^*E_y>)$$
 $$=-I_1\sin2\alpha_1\sin2\beta_1-I_2\sin2\alpha_2\sin\beta_2$$
 But in this formulation,$-I\sin2\alpha\sin\beta=V$,
 \begin{equation}
     \boxed{V=V_1+V_2}
 \end{equation}
 The position angle is given by,
 $$\tan{2\psi}=\frac{U}{Q}$$
 $$=\frac{I_1\sin2\alpha_1\cos\beta_1+I_2\sin2\alpha_2\cos\beta_2}{I_1\cos2\alpha_1+I_2\cos2\alpha_2}$$
 $$=\frac{I_1\tan{2\alpha_1}\cos\beta_1\cos2\alpha_1+I_2\tan2\alpha_2\cos2\alpha_2\cos\beta_2}{I_1\cos2\alpha_1+I_2\cos2\alpha_2}$$
 $$=\frac{\tan(2\psi_1)\cos2\alpha_1+I_2\tan(2\psi_2)\cos2\alpha_2}{I_1\cos2\alpha_1+I_2\cos2\alpha_2}$$
 but, $\psi_2=\psi_1+\pi/2$ implies $\tan(2\psi_1)=\tan(2\psi_2)$
$$\tan(2\psi)=\tan(2\psi_1)\frac{I_1\cos2\alpha_1+I_2\cos2\alpha_2}{I_1\cos2\alpha_1+I_2\cos2\alpha_2}$$
$$\implies \tan(2\psi)=\tan(2\psi_1)$$
\begin{equation}
    \boxed{\psi=\psi_1+\frac{n\pi}{2}}
\end{equation}
This implies that only orthogonal jumps are allowed. If there are two orthogonal incoherent waves, and one has decreasing amplitude and another has increasing amplitude with time, then Stokes $Q$ and $U$ will slowly go to zero ( will be zero when the amplitude of Q and U are same from these two waves) making linear polarization zero and then they will simultaneously change their sign (Orthogonal modes have opposite signs of (Q,U), due to $\pi$ rotation in Q-U plane).  For the first part (before the linear goes to zero) PPA will remain in one mode (mode of the dominant wave) and after the Q and U have changed their signs, It will undergo an orthogonal jump and remain there (until another mode dominates).

\bsp	
\label{lastpage}
\end{document}